\documentclass[useAMS,usenatbib]{mn2e}

\usepackage{graphics}

\title[Possible arcminute-separation gravitational lensed QSOs in the
2dF QSO survey]{Possible arcminute-separation gravitational lensed QSOs in the
2dF QSO survey}
\author[L. Miller et al.]
{L. Miller$^{1}$,\thanks{http://www-astro.physics.ox.ac.uk/${\tiny \sim}$lam} 
A.M. Lopes$^{1}$,
R.J. Smith$^{2}$,
S.M. Croom$^{3}$,
B.J. Boyle$^{3}$,
T. Shanks$^{4}$,\newauthor
P. Outram$^{4}$.
\\
$^{1}$Department of Physics, Oxford University,
Denys Wilkinson Building, Keble Road, Oxford OX1 3RH, U.K.\\
$^{2}$Astrophysics Research Institute, Liverpool John Moores University, 
Twelve Quays House, Egerton Wharf, Birkenhead CH41 1LD, U.K.\\
$^{3}$Anglo-Australian Observatory, PO Box 296, Epping, NSW 2121, Australia.\\
$^{4}$Department of Physics, Durham University,
Science Laboratories, South Road, Durham DH1 3LE, U.K.
}

\begin{document}

\date{\today}

\pagerange{\pageref{firstpage}--\pageref{lastpage}} \pubyear{2002}

\maketitle

\label{firstpage}

\begin{abstract}
We report the possible discovery of multiple gravitationally-lensed
images of QSOs with angular separations on arcminute scales.  The
QSOs were selected from the completed 2dF QSO Survey
as having redshifts and optical colours which are statistically consistent.
In this paper we present higher-quality optical spectra of the candidates
and discuss the likelihood of their genuinely being systems lensed
by massive clusters of galaxies.  From a comparison of the
spectra it appears that up to six pairs of QSOs may be lensed 
multiple images, although the true number may be less than that and 
further observations should be undertaken to amass more
evidence and to detect the lensing clusters.
Two of the candidates may be associated with low redshift clusters of
galaxies. 
\end{abstract}


\section{Introduction}

Strong gravitational lensing of background objects by massive clusters
of galaxies is expected to produce multiple images with wide
separations, $> 30''$, and the multiple images of background lensed
galaxies have long been known with radii of giant arc
systems up to $70''$
(e.g. \citealt{kneib96, sahu98, smail95}).  However, 
despite a number of searches,
primarily in large-area radio surveys 
(e.g. \citealt{maoz93, maoz97, ofek01, ofek02, phillips01a, phillips01b}),
widely-separated multiple images of QSOs have not yet been 
found.  This is
attributable to the low probability of strong lensing along any given
line of sight coupled with the relatively low numbers of known QSOs.  The
widest separation QSO lensed systems have separations $\sim 7''$ 
\citep{CASTLES}.  Other wide-separation multiple QSO systems found to date
are more likely to be ``binary QSOs'' (that is, a pair of active black
holes cohabiting the same parent host galaxy) rather than multiple
images of the same QSO \citep{koch99}.  

We have searched a catalogue of 22,163 QSOs that comprises the 
recently completed 2dF QSO survey (``the 2QZ survey'') for candidate 
wide-separation lensed QSO systems and have obtained higher signal-to-noise
spectra in an attempt to eliminate obvious non-lensed systems and to
identify systems that are likely to be genuine lensed systems.  After
an incomplete survey of the most likely candidates, there remain 
six systems which,
with varying degrees of confidence, are likely to be lensed.

\section[]{The selection of candidate lensed systems from the 2QZ survey}

The 2QZ survey covers 740 deg$^{2}$ in two strips of sky, one centred on the
South Galactic Pole region and the other on the celestial equator in the region
of the North Galactic Pole.  The survey photometry is described by
\citet{smith02} and the initial release of the first 10,000 QSOs 
is described by \citet{croom01}.  In constructing the 2QZ survey,
candidate QSOs were first selected from
$u$, $b_J$ and $r$ photographic photometry of morphologically-stellar objects, 
and objects whose colours lay
bluewards of the main stellar locus were selected for low resolution
spectroscopy at the two-degree field (2dF) facility at the Anglo-Australian
Telescope.  

The initial spectra obtained were of sufficient quality that reliable
identifications of QSOs and measurements of their redshifts could be made,
with an r.m.s. error in redshift measurement of typically 0.0035 
\citep{croom01}, although it should be noted that the redshift error for
an individual object depends on its brightness and emission-line equivalent
widths as well as observing conditions.  The absolute accuracy of the 
photometry in each band is estimated as $\sim 0.17$ magnitudes at the
faint survey limit
\citep{smith02}.  The relative photometric accuracy for objects measured
close together on the same photographic plate is better than this,
however: \citet{mitchell89} found that, for objects more than two magnitudes
brighter than the completeness limit of the photographic plate, relative
photometry accurate to $\sim 0.05$ magnitudes was typical, with that value
rising rapidly to values $> 0.1$ magnitudes as the plate limit is approached.

The resulting catalogue of photographic photometry and spectroscopically
confirmed QSOs was then searched for pairs or multiple systems whose
colours were consistent given the large photometric uncertainties and 
whose redshifts were consistent given the typical redshift uncertainty.
To determine the colour consistency, it was assumed that two QSOs, if
lensed, should have the same $u$, $b_J$ and $r$ magnitudes apart from a
magnitude offset arising from differential magnification of the images
in a lensed system, and thus assuming negligible effects due to
variations in QSO colours with time (see section~\ref{timedelay}).  
The statistic used as the measure of colour consistency was
$$
\sigma_{\scriptscriptstyle UBR}^{2} = \sum_{i=1}^{3}(\Delta m_{i}-\overline{\Delta m_{i}})^{2},
$$
where $\Delta m_{i}$ is the difference in apparent magnitude of the
two QSOs measured in 
waveband $i$ and $\overline{\Delta m_{i}}$ is the average 
magnitude difference for that QSO pair.
Redshift consistency was
measured by the difference, $|\Delta z|_{\rm 2QZ}$ in 2QZ catalogue redshift of two QSOs.  
Candidate multiple systems were initially selected if they satisfied the
criteria $\sigma_{\scriptscriptstyle UBR} < 0.2$, $|\Delta z|_{\rm 2QZ} < 0.01$ and angular
separation $\Delta \theta < 6'$.  The colour criterion is close to the
expected scatter in this statistic, given the photometric errors at the
survey limit, and hence we expect the initial colour selection to select 
only $\sim 50$\% of the lensed systems.  The redshift
criterion is a factor two larger than the expected scatter in 
$|\Delta z|_{\rm 2QZ}$ and this criteria should allow the majority of lensed 
systems to be selected.  The stringent colour criterion was chosen in
this initial study in order to maximise the probability of finding
lensed systems, and hence further observations should be carried out to
establish the true completeness of the selected sample.

A further check was made on the consistency of their radio properties by 
searching for detections in the NVSS \citep{nvss}.  One candidate pair with
clearly discrepant radio detections was removed from the sample of
candidates.  The NVSS properties of two remaining pairs
are discussed further in section 3.

Candidates with $18.25 < b_J < 20.85$ were selected 
from 12307 QSOs in the SGP region and from
9856 QSOs in the NGP region of the 2QZ survey, and 38 pairs of QSOs
in total met these criteria.  Of the QSOs with confirmed redshifts, there
were no candidate systems selected with more than two members, although
some of the pairs do have additional $ubr$-selected companions for which 
2QZ spectroscopic observations either were not carried out or did not
lead to a positive identification.

Our expectation is that the majority of the selected pairs will
not prove to be lensed systems.  Test samples were also constructed
in attempt to assess the possible contamination.  First, 
QSOs pairs that are more widely separated, but still within one degree
of each other and measured on the same photographic plates, were selected
with $|\Delta z|_{\rm 2QZ} < 0.02$ and the distribution of $\sigma_{\scriptscriptstyle UBR}$ 
was compared with the $\sigma_{\scriptscriptstyle UBR}$ distribution of the candidate 
pairs. No statistically-significant difference was found, although
this is probably a reflection of the large errors in $\sigma_{\scriptscriptstyle UBR}$
arising from the large photographic photometry errors.  A second test
sample comprising QSO pairs with $\sigma_{\scriptscriptstyle UBR} > 0.5$ and 
$|\Delta z|_{\rm 2QZ} < 0.01$ was selected and the distribution of angular
separations, normalised by the total number of pairs selected with
$\Delta \theta < 10'$, was compared with the candidate lens sample
selected to have $\sigma_{\scriptscriptstyle UBR} < 0.2$.  
This time there was an excess of pairs on arcminute scales
in the sample of candidate lensed systems 
consistent with a minority of the pairs
being lensed systems, but because of the small numbers of pairs selected
the test was not statistically significant.  Hence at this stage there is
no evidence either for or against the hypothesis that the selected sample 
contains a subset of lensed systems.  Further evidence needs to be collected
on all members of the sample in order to establish its completeness and
contamination.

A key part of that evidence is the follow-up spectroscopy described in
the next section.  Because of limitations of 
observing time and observing conditions, only a subset of the candidates
have had such observations, and only that subset is described in the
remainder of this paper.  For prioritising the observations, the pairs
were ranked in $|\Delta z|$ and also in $\sigma_{\scriptscriptstyle UBR}$:
a composite rank was produced by multiplying these ranks and the top
candidates in each region were observed.
In the SGP region, 10 pairs were observed, but 
the spectroscopic follow-up is particularly
sparse in the NGP region, where only for one pair have spectra been obtained of
sufficient quality to allow some assessment of whether or not it may
be lensed.  At this stage the sample reported here cannot be regarded as
either being complete or free from contamination by non-lensed pairs of QSOs.
We postpone any discussion of the completeness of the sample to later
work.  We anticipate that in future the candidate QSO pairs in the NGP
region will also have the benefit of more accurate, five-band, colours
from the Sloan Digital Sky Survey \citep{stoughton02}, which should lead
both to samples of candidates with lower contamination and with better
quantified photometric uncertainties.  

In the following sections each pair is designated by the IAU naming
convention applied to the mean position of the pair, in the FK5
(equinox J2000) coordinate system.

\begin{table*}
\begin{minipage}{170mm}
\caption{Observed QSO pairs. The columns are: mean right ascension of the pair
(equinox 2000), mean declination, mean 2QZ survey redshift, colour difference statistic,
2QZ redshift difference and angular separation,
followed by the right ascension, declination and 2QZ catalogue apparent b
magnitude for each member of the pair.}
\begin{tabular}{|c@{ }c@{ }c@{ }c@{ }c@{ }r@{ }c@{ }c@{ }c@{ }c@{ }c@{ }c|}
\hline
$\overline{\rm R.A.}$ (J2000)& $\overline{\rm dec}$ (J2000) & $\overline{\rm z}$ & $\sigma_{\scriptscriptstyle UBR}$ & $|\Delta z|_{\rm 2QZ}$ & $\Delta\theta''$ & R.A. (J2000) & dec (J2000) & b & R.A. (J2000) & dec (J2000) & b \\
\hline
14 35 07.4 & $+$00 08 53 & 2.378 & 0.03 & .0086 &  33 & 14 35 08.32 & $+$00 08 44.4 & 20.13 & 14 35 06.42 & $+$00 09 01.5 & 20.03 \\
01 50 55.6 & $-$29 21 57 & 1.556 & 0.09 & .0011 & 200 & 01 50 50.10 & $-$29 23 06.3 & 19.77 & 01 51 01.10 & $-$29 20 47.5 & 20.34 \\
01 35 31.2 & $-$30 21 54 & 1.683 & 0.14 & .0009 &  46 & 01 35 31.01 & $-$30 22 17.1 & 20.69 & 01 35 31.49 & $-$30 21 31.9 & 20.45 \\
02 15 55.8 & $-$29 06 21 & 2.264 & 0.07 & .0063 & 347 & 02 15 48.09 & $-$29 08 42.0 & 20.08 & 02 16 03.54 & $-$29 03 59.8 & 20.79 \\
02 23 00.7 & $-$29 49 05 & 1.416 & 0.02 & .0018 & 236 & 02 22 59.28 & $-$29 51 01.5 & 19.64 & 02 23 02.03 & $-$29 47 08.0 & 20.50 \\
22 28 23.4 & $-$28 57 35 & 2.453 & 0.17 & .0011 & 270 & 22 28 13.72 & $-$28 58 22.1 & 20.64 & 22 28 33.07 & $-$28 56 48.9 & 19.94 \\
22 35 20.5 & $-$27 17 18 & 1.810 & 0.07 & .0049 & 316 & 22 35 16.20 & $-$27 19 45.4 & 18.50 & 22 35 24.82 & $-$27 14 50.5 & 20.16 \\
22 10 05.1 & $-$32 04 00 & 2.119 & 0.11 & .0061 & 330 & 22 09 56.21 & $-$32 06 00.3 & 20.81 & 22 10 14.03 & $-$32 02 00.0 & 20.71 \\
22 03 09.4 & $-$31 33 49 & 1.643 & 0.19 & .0015 &  89 & 22 03 11.80 & $-$31 34 21.6 & 19.75 & 22 03 07.09 & $-$31 33 16.5 & 19.43 \\
22 50 32.1 & $-$31 05 00 & 0.637 & 0.19 & .0059 & 242 & 22 50 22.71 & $-$31 04 45.4 & 18.40 & 22 50 41.45 & $-$31 05 14.4 & 20.17 \\
23 29 07.8 & $-$29 57 00 & 1.841 & 0.12 & .0002 &  67 & 23 29 06.87 & $-$29 57 31.9 & 20.28 & 23 29 08.69 & $-$29 56 28.7 & 19.70 \\
\hline
\end{tabular}
\end{minipage}
\label{qsopairs}
\end{table*}

\section[]{Further optical spectroscopy}

Spectra were obtained of the candidate pairs 
at the William Herschel 4.2-m. telescope (WHT) on 2002 February 5-6 
and at the Anglo-Australian 3.9-m. 
telescope (AAT) on 2002 September 6-10.
At the WHT, the ISIS long-slit spectrograph with 300B grating and EEV detector
was used with on-chip binning to provide spectra from the blue arm with
1.7\AA~pixel$^{-1}$ dispersion.  Observing conditions were poor, with
seeing of about $2''$ for the spectra reported here, resulting in 
spectra with resolution about 8\AA.  Data were also obtained with the ISIS 
red arm but were of insufficient quality to be useful.
Observations at the AAT were carried out in generally good conditions with 
seeing in the range $0.7 - 1.6''$.  The 300B grating on the RGO long-slit
spectrograph with EEV detector provided a dispersion 
of 1.6\AA~pixel$^{-1}$ and resolution
of typically 6\AA (depending on the choice of slit width, which was varied
according to the seeing at the time).

The typical integration time on each QSO varied between one hour and
four hours depending on observing conditions and brightness of the QSO. 
In most cases the QSOs in each pair were sufficiently far apart that 
they could not be observed simultaneously.  Instead, integrations
on each QSO were split into intervals of 30 minutes duration, and no
more than three integrations were carried out sequentially on one QSO
without then observing the other member of the pair.  By this process we
could be sure that the spectra were taken under sufficiently similar
conditions that any differences between them are likely to be genuine.
For those QSOs observed
in this way the slit position angle was arranged to be at the parallactic angle.
In two cases, J1435$+$0008 and J0135$-$3021,
the pair members were close together and were observed simultaneously with
the slit oriented at the appropriate angle.  
In the case of J1435$+$0008 the pair could be observed at hour angles such
that the slit position angle was within 20 degrees of the parallactic angle.
In the case of J0135$-$3021 this was not possible, but the pair could
be observed within 30 degrees of the zenith where the 
effects of differential refraction should be small.

\begin{figure*}
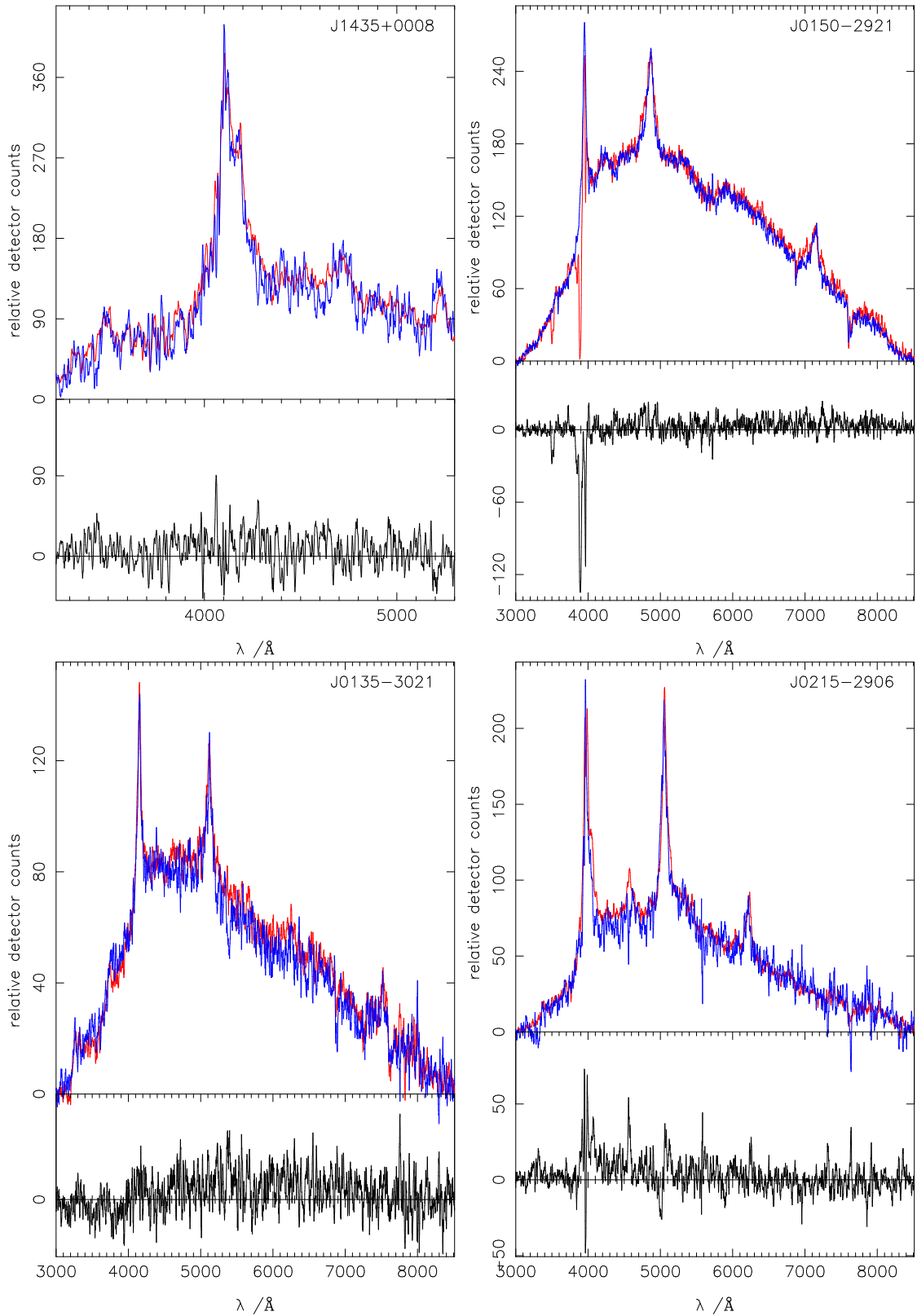

\begin{minipage}{168mm}
\resizebox{74mm}{!}{
\rotatebox{0}{
\includegraphics{J1435+0008_s2.ps}}}
\resizebox{74mm}{!}{
\rotatebox{0}{
\includegraphics{J0150-2921_s3.ps}}}

\resizebox{74mm}{!}{
\rotatebox{0}{
\includegraphics{J0135-3021_2nd_s4.ps}}}
\resizebox{74mm}{!}{
\rotatebox{0}{
\includegraphics{J0215-2906_s4.ps}}}

\caption{
Scaled and superimposed spectra of the candidate
lensed systems. The differences between the scaled spectra 
are also shown. To help distinguish
genuine features from noise, all spectra are shown without any
relative flux calibration applied.
Atmospheric absorption features are still present in the uncalibrated spectra.
The flux scale is the number of detector counts observed per pixel in
individual exposures (typically of duration 1800~s) of one of the QSOs,
with the count values of the second QSO scaled to match.
{\em upper left} WHT ISIS blue-arm spectra of 
J1435$+$0008 smoothed to a resolution of 12\AA.  
{\em upper right} AAT RGO spectrograph spectra of J0150$-$2921
smoothed to a resolution of 12\AA.
{\em lower left} AAT RGO spectrograph spectra of J0135$-$3021
smoothed to a resolution of 15\AA.
{\em lower right} AAT RGO spectrograph spectra of J0215$-$2906
smoothed to a resolution of 15\AA.
\label{fig1}
}
\end{minipage}
\end{figure*}

\setcounter{figure}{0}
\begin{figure*}
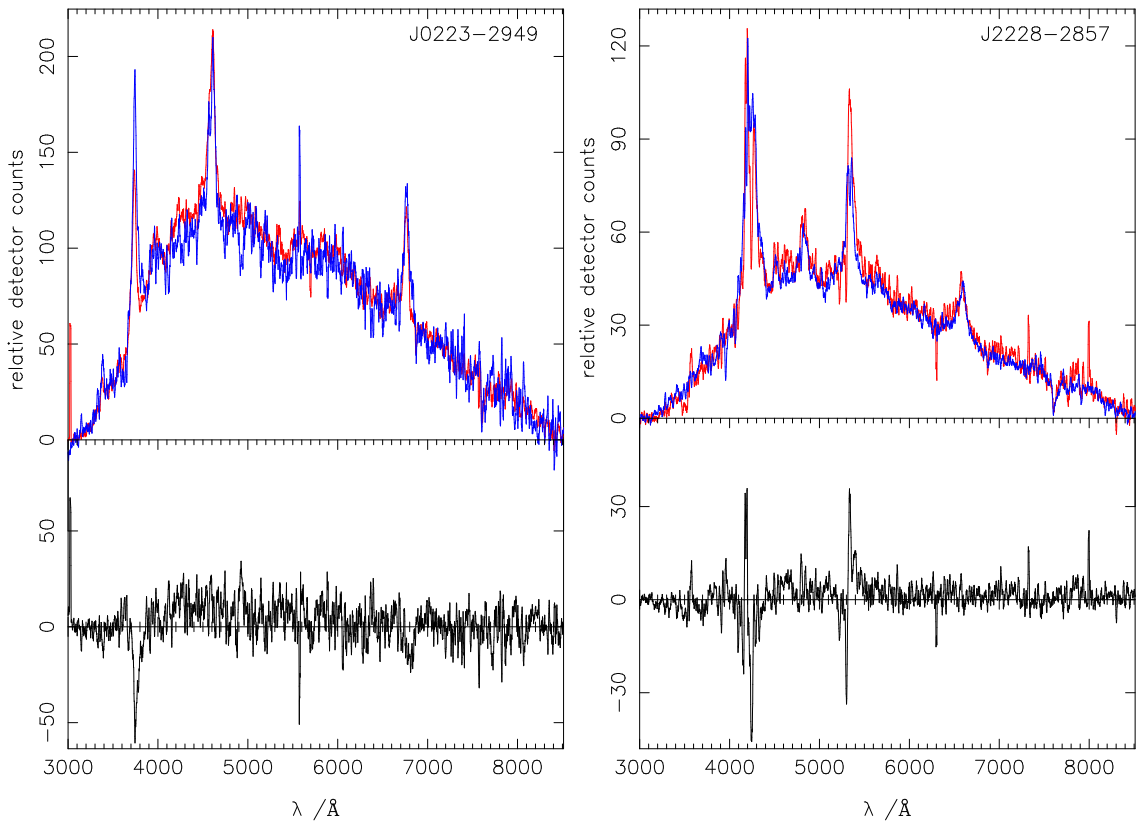

\begin{minipage}{170mm}
\resizebox{75mm}{!}{
\rotatebox{0}{
\includegraphics{J0223-2949_s4.ps}}}
\resizebox{75mm}{!}{
\rotatebox{0}{
\includegraphics{J2228-2857_s4.ps}}}
\caption{
(continued):
{\em left} AAT RGO spectrograph spectra of J0223$-$2949
smoothed to a resolution of 15\AA.
{\em right} AAT RGO spectrograph spectra of J2228$-$2857
smoothed to a resolution of 15\AA.
}
\end{minipage}
\end{figure*}

\setcounter{figure}{0}
\begin{figure*}
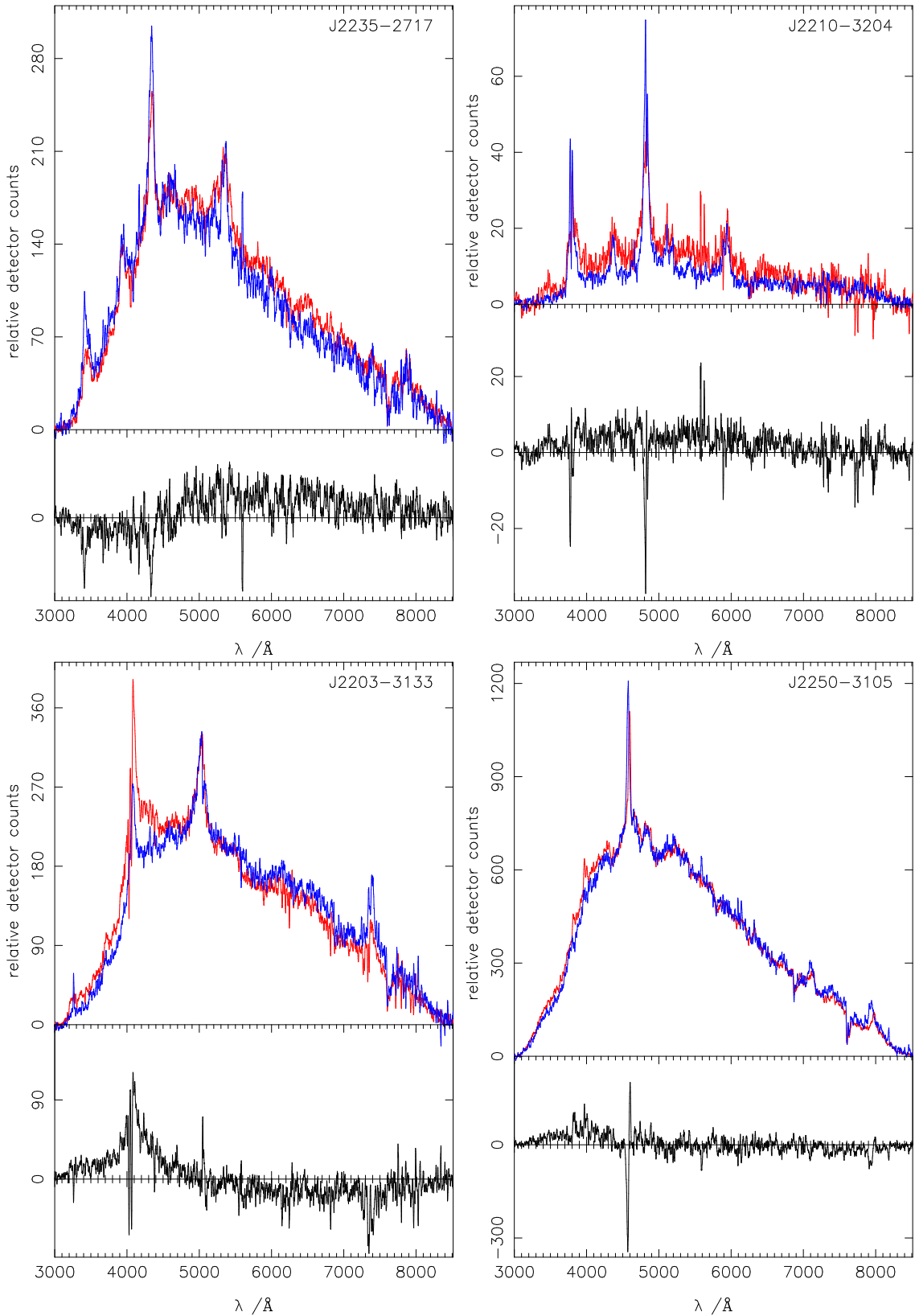

\begin{minipage}{170mm}
\resizebox{75mm}{!}{
\rotatebox{0}{
\includegraphics{J2235-2717_s4.ps}}}
\resizebox{75mm}{!}{
\rotatebox{0}{
\includegraphics{J2210-3204_s3.ps}}}

\resizebox{75mm}{!}{
\rotatebox{0}{
\includegraphics{J2203-3133_s3.ps}}}
\resizebox{75mm}{!}{
\rotatebox{0}{
\includegraphics{J2250-3105_s3.ps}}}

\caption{
(continued):
{\em upper left} AAT RGO spectrograph spectra of J2235$-$2717
smoothed to a resolution of 15\AA.
{\em upper right} AAT RGO spectrograph spectra of J2210$-$3204
smoothed to a resolution of 12\AA.
{\em lower left} AAT RGO spectrograph spectra of J2203$-$3133
smoothed to a resolution of 12\AA.
{\em lower right} AAT RGO spectrograph spectra of J2250$-$3105
smoothed to a resolution of 12\AA.
}
\end{minipage}
\end{figure*}

Spectra of all the pairs observed are presented in Fig~\ref{fig1}.  
In order to improve the signal-to-noise ratio per pixel the spectra
have been smoothed with a top-hat function to a resolution specified
in the figure caption for each pair. Each
panel shows spectra of the two members of each pair superimposed.
The spectra have been scaled by the total value of the observed
(uncalibrated) count rate so that they may be compared. Each panel also
shows the difference between the scaled spectra, for comparison.
The spectra have been bias-frame subtracted and flatfielded in order to
remove instrumental features from the spectra, but have
not had any flux calibration applied, in order to facilitate
evaluation of any differences.  The noise in the spectra
is dominated by photon noise from the sky background, and the noise
in each wavelength pixel is more uniform with wavelength in uncalibrated spectra.
Absorption features arising from the terrestrial atmosphere remain in
the spectra, most prominently at $\lambda 7600$\AA, and should be ignored
when comparing spectra.
The ratios of the total count rate between the spectra are given in
Table~\ref{ratios}.

One expectation of the ``lensed QSO'' hypothesis is that the two
images should have the same redshift.  As the pairs were selected to
have 2QZ survey redshift differences $|\Delta z|_{\rm 2QZ} < 0.01$ we
would, in the absence of QSO clustering, expect the distribution of
WHT or AAT redshift differences to be uniformly distributed within
that range.  QSO clustering has a scale length in redshift of about
0.003 at $z \sim 1.5$ \citep{croom01b}, and if the QSOs in each pair
are individual objects we would expect them also to have redshift
differences arising from their peculiar velocities, so the ``separate
QSOs'' hypothesis predicts that even in the presence of QSO clustering
there should still be a broad distribution of redshift differences
when measured from the WHT or AAT spectra.

\setcounter{figure}{0}
\begin{figure}
\resizebox{75mm}{!}{
\rotatebox{0}{
\includegraphics{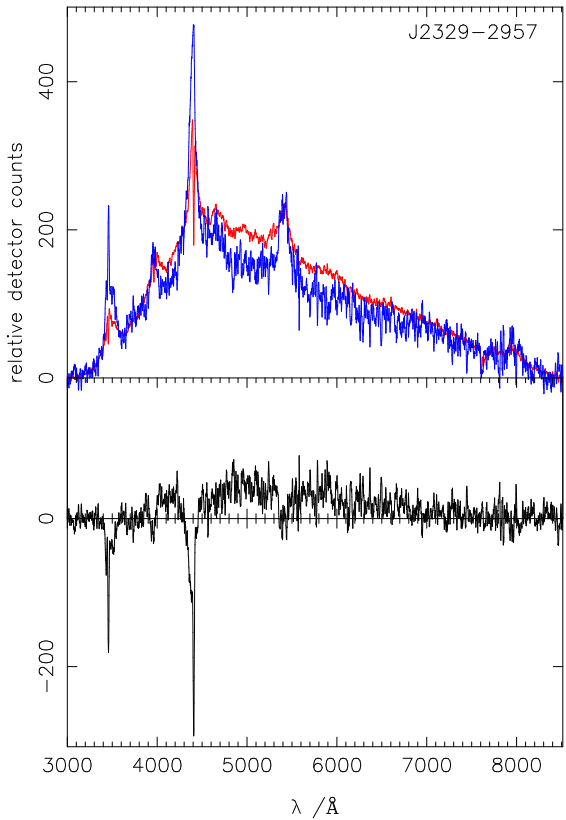}}}
\caption{
(continued): 
Uncalibrated AAT RGO spectrograph spectra of J2329$-$2957
smoothed to a resolution of 12\AA.
}
\end{figure}

\begin{figure}
\resizebox{75mm}{!}{
\rotatebox{-90}{
\includegraphics{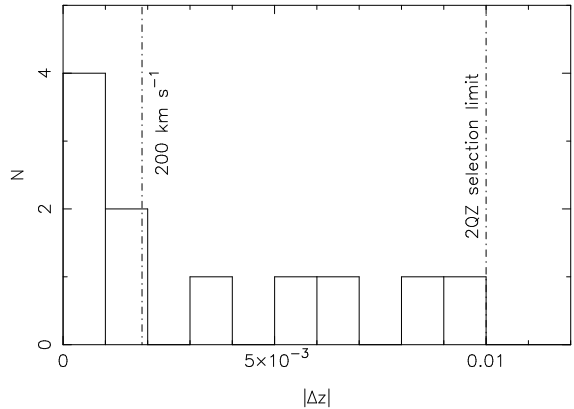}}}
\caption{
The histogram of WHT/AAT redshift differences, $|\Delta z|$, listed
in Table~\ref{ratios}.  The original selection limit based on 2QZ
redshifts is shown, as is the redshift difference expected for a
pair at the mean sample redshift of $z=1.8$ with a relative 
peculiar velocity difference of 200~km~s$^{-1}$ 
\label{zdiff}
}
\end{figure}

To test this, the redshift differences between each pair of WHT or AAT
spectra have been measured by cross-correlation, about subtracting the
continuum from the uncalibrated spectra and avoiding the atmospheric
absorption feature at $\lambda 7600$\AA.  Uncertainties in the
redshift differences have been estimated assuming the noise in the
spectra is dominated by uncorrelated photon shot noise in each pixel.
A Monte-Carlo method was adopted in which Gaussian noise of the same
amplitude as the noise in each spectrum was added to smoothed versions
of each spectrum to create a number of synthetic spectra with the same
spectral features and noise properties as the actual data. The
synthetic spectra were then cross-correlated with each other and the
quoted uncertainty in the redshift difference is the rms of the values
derived from the synthetic spectra.  These uncertainties are lower
limits to the true uncertainty as they contain no allowance for
systematic differences and noise that may be correlated between
pixels.  However, the spread in values of the measured redshift
differences is consistent with these uncertainty estimates, indicating
that photon shot noise is the dominant source of uncertainty.
Remarkably, we find six of the pairs to have velocity differences less
than 200~km~s$^{-1}$ (Table~\ref{ratios} and Fig.~\ref{zdiff}).  This
is a strong indication that a number of these pairs are lensed
systems.  All separations for these pairs are sufficiently large that,
if they were two different QSOs, the linear separations would be greater
than 1\,Mpc and they must occupy separate host galaxies.  QSO clustering
would cause a peak in the redshift distribution at $|\Delta z| = 0$,
but broadened not only by the measurement error but also by the
relative peculiar velocities of the QSO host galaxies.  Peculiar
velocities at high redshift are expected to be lower than in
low-redshift galaxy surveys, but simulations still predict pairwise velocity
dispersions $\sigma \sim 350$\,km\,s$^{-1}$ \citep{zhao02}, significantly
larger than would be inferred from Fig.~\ref{zdiff}.

Although the redshift difference is a quantitative measure of similarity 
between
the spectra, we should also look for other differences in the spectra.
We now give a brief evaluation of the comparison of the spectra of each pair.  
The discussion is ordered by an approximate degree of confidence in the
likelihood that the pair is a lensed QSO.  This evaluation is based not
only on redshift difference but also on
similarity of continuum shapes, emission line equivalent widths, line ratios,
velocity widths and profiles. This comparison is necessarily somewhat subjective.  
It is never possible to prove that a pair of
spectra are multiple images of the same lensed QSO, and in this case we have
to rely on building up a body of evidence which can then be interpreted
as either supporting the lensed hypothesis or providing evidence against
it.  The spectra we have obtained only cover the range of accessible
optical wavelengths and vary widely in their signal-to-noise.  The spectra
of a QSO pair
viewed with high signal-to-noise may appear to be discrepant,
such as J2250$-$3105, whereas such a discrepancy may not have been apparent
in lower signal-to-noise spectra.  The only reliable solution to this
effect is to obtain higher quality spectra and other evidence
for those pairs with poor signal-to-noise.  With these
factors in mind, any evaluation at this stage cannot be regarded as conclusive,
but nonetheless should help us to identify those QSOs that are most
likely to be lensed, that can then be the subject of further observations
to amass more evidence.  The candidates we have observed were selected
{\em a priori} to have similar optical colours (albeit with a large
uncertainty) and redshifts, so we should not be too surprised
to find that almost all of the observed pairs do indeed have very similar
overall continuum shapes in the observations presented here.
Instead, we must appeal to the wide range of detail in QSO spectra which
we would not expect to match when comparing the spectra of independent
QSOs.

\begin{table}
\caption{Observed redshift differences and flux ratios between the 
WHT/AAT spectra. The flux ratio is defined as the
ratio of the flux from the second listed QSO in Table~1 
to the flux from the first listed QSO of each pair.}
\begin{tabular}{|ccc|}
\hline
Pair & redshift difference & spectrum flux ratio \\
\hline
J1435$+$0008 & $+0.0004 \pm 0.0016$ & 0.40 \\
J0150$-$2921 & $-0.0010 \pm 0.0007$ & 0.72 \\
J0135$-$3021 & $+0.0051 \pm 0.0014$ & 0.71 \\
J0215$-$2906 & $-0.0001 \pm 0.0011$ & 0.27 \\
J0223$-$2949 & $+0.0015 \pm 0.0010$ & 0.32 \\
J2228$-$2857 & $+0.0037 \pm 0.0016$ & 2.13 \\
J2235$-$2717 & $+0.0000 \pm 0.0019$ & 0.28 \\
J2210$-$3204 & $+0.0017 \pm 0.0016$ & 2.00 \\
J2203$-$3133 & $-0.0100 \pm 0.0015$ & 0.67 \\
J2250$-$3105 & $-0.0063 \pm 0.0005$ & 0.27 \\
J2329$-$2957 & $-0.0089 \pm 0.0022$ & 0.19 \\
\hline
\end{tabular}
\label{ratios}
\end{table}

\subsection{J1435$+$0008}

This pair was the only one presented here that was observed at the WHT. 
Other pairs were attempted but only inadequate spectra were obtained.
Observing conditions were poor, with
poor atmospheric transparency and poor seeing.  The ISIS red arm spectra
contain too little signal, and we show here a comparison of the blue-arm 
spectra only.  The dichroic between the red and blue arms of ISIS causes a sharp
fall in response at $\lambda > 5000$\AA, with some oscillations in that
response in this wavelength range.
This is a good lensed QSO candidate, with a small redshift difference
and an unusual
ratio of Ly-$\alpha$ and NV which is common to both spectra. 

This QSO pair also appears in the Sloan Digital Sky Survey (SDSS) 
Early Data Release \citep{stoughton02}.  Table~\ref{EDR} gives the
SDSS photometry for this pair, together with the differences in 
apparent magnitudes and the associated error.  It can be seen that 
the colours of the two QSOs are consistent with being the same to
substantially higher accuracy than was possible with the 2QZ photographic
photometry, although there does seem to be some some evidence for a 
systematic difference in colour with waveband.  This may be taken as
evidence against the system being lensed, or it may be accounted for
either by spectral variability or 
by differential extinction between the light paths.  

We also note the possibility of an elegant test of the lensed-QSO hypothesis:
if these are images of the same QSO then, at redshifts near the QSO, the
Ly-$\alpha$ forest absorption lines should be identical in both spectra.
If the pair were not a single lensed QSO, then the QSO separation would
be $\sim 1$\,Mpc and the Ly-$\alpha$ lines would not be expected to be
common between the QSOs: sight lines that are separated
on such scales when comparing unrelated QSOs show significant
differences in their Ly-$\alpha$ forest absorption \citep{dodorico}.  
Forest lines at redshifts significantly different
from the QSO may nonetheless differ between the images in either hypothesis
because the light rays for each image would have diverged to a significant 
extent.  In the case of J1435$+$0008 the ultra-violet part of the spectrum
is noisy, but nonetheless it does appear to have possible absorption
features in common between the spectra.

\subsection{J0150$-$2921}

This is possibly one of the most enigmatic pairs of spectra obtained.  The
spectra have good signal-to-noise ratio and appear identical in every respect
apart from one gross feature:  the presence of a broad absorption line system
in one of the spectra.  For reasons described in section \ref{timedelay}
we do not believe that this should be taken
as evidence against the system being a lensed
QSO, and because of the otherwise striking similarity between the spectra 
in these observations with good signal-to-noise
we regard this as being one of the systems most likely to be a
single lensed QSO.

This system also has a number of bright galaxies in the vicinity, of which
three have had redshifts measured by the 2dF Galaxy Redshift Survey
\citep{colless03}.  
The positions and redshifts of the two QSOs and of the
three galaxies are shown in Fig.~\ref{fchart}.  This system is discussed
further in section 5.

We also note here that although neither of the QSOs is detected in the NVSS
survey there are four detected sources of radio emission extending over
some arcminutes approximately symmetrically placed with respect to the
centre of the system, and approximately perpendicular to it.  It may be
that the radio emission is the substructure of a single extended double radio source associated
with one of the galaxies in the field.

\subsection{J0135$-$3021}

There are some small differences visible in the ultraviolet parts of the spectra
of this pair 
which lead to the somewhat high value of redshift difference given in 
Table~\ref{ratios}.  However,
these spectra were taken simultaneously, with the slit rotated to the
position angle defined by the coordinates of the two QSOs, and the
observations presented here were made with the pair within about thirty degrees
of the zenith.  Slight errors in aligning the slit with the pair could
result in a small differential loss of light owing to atmospheric refraction
which may contribute to the small differences at $\lambda < 4500$\AA.  At
this stage we still regard this pair as possibly being a lensed QSO, although
further observations are needed.

\subsection{J0215$-$2906}

This pair has very similar spectra with a redshift difference of zero to
high accuracy.  There are some residual features
present in the difference of the scaled spectra which may argue against the
lensed QSO hypothesis but which may simply be observational noise or
may reflect low-level emission variations (see also section \ref{timedelay}).

\subsection{J0223$-$2949}

This pair has spectra identical within the noise except in the CIV line, where
there is a residual difference whose maximum is about one-third the 
maximum of the emission-line.  The other emission lines and the detailed
continuum shape are extremely similar, however, and the redshift difference
is consistent with zero at the 2-sigma level.  We regard this as being
a possible lensed system that requires further observations.

This system also has a number of bright galaxies in the vicinity
(Fig.~\ref{fchart}), of which
ten have had redshifts measured by the 2dF Galaxy Redshift Survey
\citep{colless03}.  
This system is discussed further in section 5.
 
There is a 8.6~mJy NVSS \citep{nvss} radio source $54''$ to the north of the 
northern QSO which is not reproduced near the southern
QSO. However, that radio source 
is coincident with the bright peculiar-shaped galaxy 
at $z=0.1147$ and it is unlikely
that the radio source is related to the QSO.

There are also weaker NVSS detections at the NVSS detection 
limit, one a 2.8~mJy source $45''$ from the southern QSO, the other
a 3.5~mJy source $24''$ from the northern QSO.  Both the reality
of these weak radio detections and the possibility of their being associated
with the QSOs needs investigation with a higher resolution, higher
sensitivity radio map.

\subsection{J2228$-$2857}

This pair has very similar spectra but does display some significant
differences in the Ly-$\alpha$/NV and CIV lines.  This could be a lensed system
if we are prepared to accept the possibility of time variations in 
emission-line strength and/or QSO intrinsic absorption features 
(see section \ref{timedelay}).

\subsection{J2235$-$2717}

Although the redshift difference for this pair is consistent with being
zero, the QSOs do have significantly different emission-line ratios, and 
we believe this pair is rather unlikely to be a lensed system.

\subsection{J2210$-$3204}

This pair exhibits significant differences in line strength between the
two spectra and is also unlikely to be a lensed QSO.

\subsection{J2203$-$3133}

This pair has quite different continuum shapes and a large redshift difference
and is unlikely to be a lensed QSO.

\subsection{J2250$-$3105}

This pair has very similar spectra, 
but there does appear to be a continuum difference between the ultraviolet parts
of the spectra and a significant redshift difference.
We consider this unlikely to be a lensed pair.

\subsection{J2329$-$2957}

This pair exhibits significant differences in emission line equivalent widths
and ratios and a significant redshift difference, making it also unlikely to be a lensed QSO.

\section[]{Time delay and the formation of broad absorption features}
\label{timedelay}

At least two of the pairs have spectra which match very well their
redshifts, continuum slopes, emission-line equivalent widths and profiles
but which do nonetheless show some significant differences between the
two QSOs.  One, J0150$-$2921, has spectra which appear identical apart
from the existence of CIV broad absorption in one QSO and not the other.
Such an observation has previously been used to argue against QSO pairs
being lensed versions of the same object \citep{green02} but the
recent discovery that broad absorption lines can form on timescales of 
ten years \citep{ma02} changes that conclusion.  

In fact, large separation lenses could have a significant time delay 
between the lensed images.  The actual time delay is strongly dependent
on the lens mass distribution and on the symmetry of the light paths
through the lens potential.  As we have so far detected
a possible lens in no more than one of these systems, 
and as the mass distributions
in any lenses are unknown, we cannot yet calculate a predicted time delay.
But in principle we should expect a delay which could be as large as
the time delay calculated from the maximum possible path length difference
in a simple geometry (and ignoring the time dilation term)
$$
\Delta t < \frac{\theta^2}{2c} \frac{D D_{\rm LENS}}{\left(D - D_{\rm LENS}\right)}
$$
where $\theta$ is the angular separation of the images, $D$ is the observer-QSO
angular diameter distance and $D_{\rm LENS}$ is the observer-lens angular
diameter distance, assuming a flat universe.  
For an angular separation of $200''$ this corresponds
to approximately 5000 years.  The closeness of the match of the spectra
of J0150$-$2921 indicates that in fact the broad absorption features have indeed
formed in one QSO within this timescale.  The available time is certainly
much greater than the 10 years formation time observed by \citet{ma02}.
One explanation for such rapid formation of absorption features is the 
ejection of remnant material from solar-mass objects being destroyed in the
vicinity of the black hole.  As the crossing time of the broad-line region
of such material is of order 10 years this seems a feasible explanation.
An alternative explanation for both the difference in our spectra and also the 
variability seen by \citet{ma02} might be that the size of a broad-absorption line gas
cloud is only of order $10^{-3}$ of the distance of the cloud from the 
central source of ultraviolet radiation.

If this conclusion is correct, it also reopens the possibility that the
$7.3''$ pair, Q2345$+$007, is a lensed QSO and not a binary QSO. 
\citet{green02} have noted that the X-ray properties differ,
but also note that one QSO shows a broad absorption line system.  If the
difference in X-ray properties is caused by X-ray absorption associated with 
the optical broad absorption line feature, as suggested by \citet{green02},
then formation of an absorption system on a timescale of order 5 years could
explain the observed results.

Two other QSO pairs exhibit some differences in their broad emission lines,
and again it is possible that these also arise from long-timescale
spectral variability. \citet{small97} have studied
spectral variability specifically with this question in mind, and
find that emission-line variations of this magnitude are common on
timescales of a few years.

Time variation may have the unwanted effect of causing the sample to be 
incomplete, if colours or spectral features vary significantly so that
either candidates are not initially selected, or if they are subsequently
rejected in a spectral comparison.  Given the potentially long time-delays
between light paths in systems of such wide separation, it is difficult 
to allow for this effect as no measurements of variability in unlensed
systems can be made on those timescales.  In the long term the best approach
would be to select candidates on the basis of observables that should be
time-independent, such as redshift difference between multiple images,
and to apply independent tests of the lensing hypothesis such as the
Ly-$\alpha$ forest comparison discussed above or looking for characteristic
features of lensing such as giant arcs associated with a QSO's host galaxy. 
In this paper we do not address this question further, preferring instead
to view this paper as providing evidence for the existence of extreme
systems which may then be used as evidence to justify more complete surveys.

\begin{figure*}
\begin{minipage}{170mm}

\resizebox{80mm}{!}{
\rotatebox{0}{
\includegraphics*{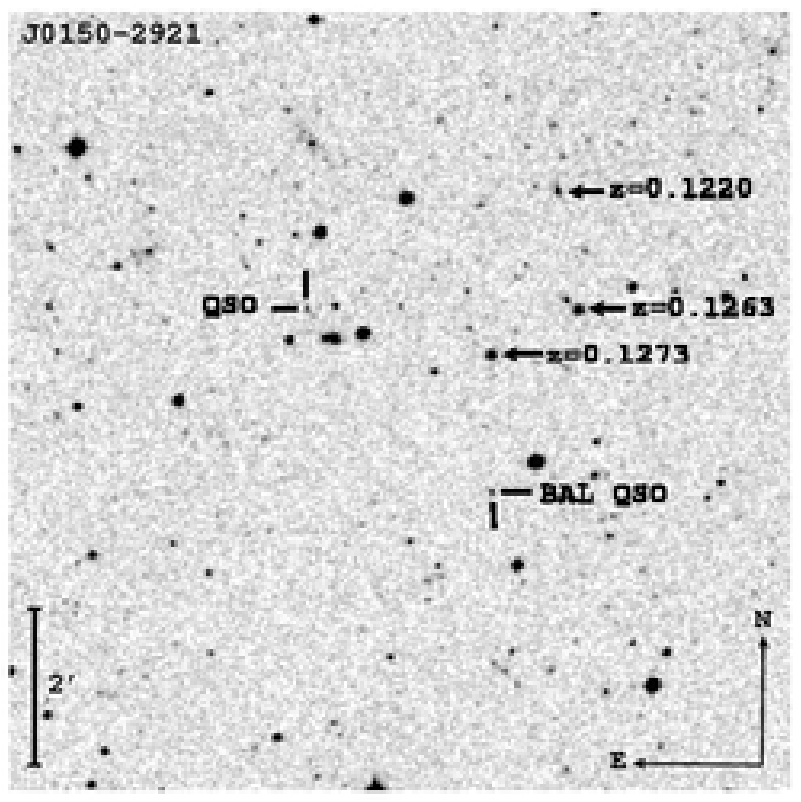}}}
\resizebox{80mm}{!}{
\rotatebox{0}{
\includegraphics*{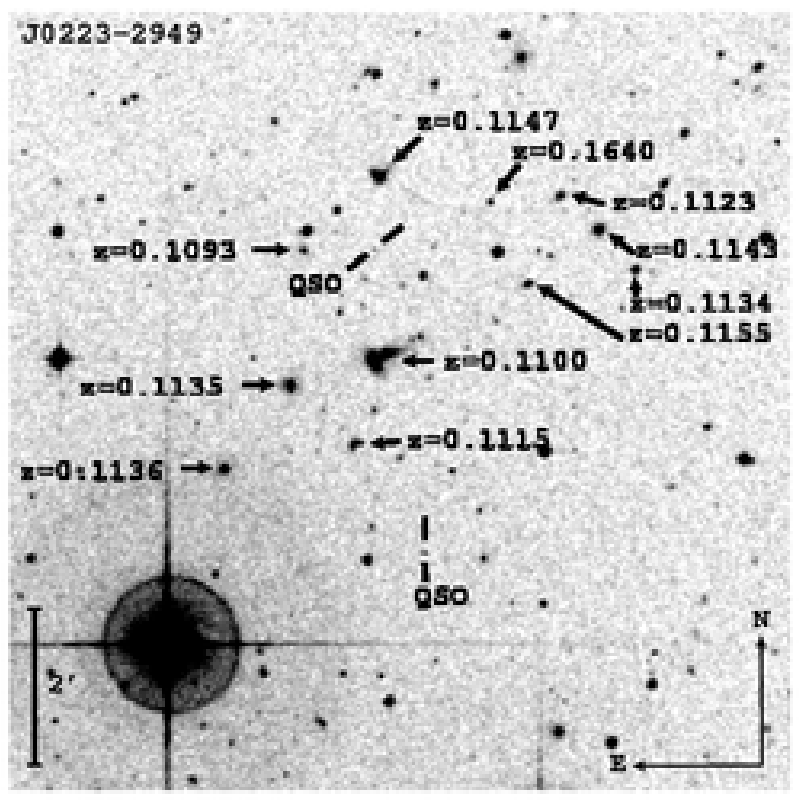}}}

\caption{ESO sky survey images of 10$'$ regions around J0150$-$2921 
and J0223$-$2949
showing the locations of both QSOs and the redshifts of the galaxies
in the vicinity measured by the 2dF Galaxy Redshift Survey.  The orientation
and image scale are also shown.
\label{fchart}
}

\end{minipage}
\end{figure*}

\begin{table}
\caption{
SDSS Early Data Release photometry for the QSO pair J1435$+$0008,
showing the apparent magnitudes in the SDSS wavebands, their
associated photometric errors, and the magnitude differences between the 
QSOs and the errors in those quantities.
}
\begin{tabular}{|lrrrrr|}
\hline
QSO & $u$ & $g$ & $r$ & $i$ & $z$ \\
\hline
J143508$+$000844 & 21.37 & 20.72 & 20.57 & 20.64 & 20.37 \\
$\pm$            & 0.09 & 0.03 & 0.03 & 0.05 & 0.13 \\
\hline
J143506$+$000901 & 20.49 & 19.92 & 19.87 & 19.98 & 19.74 \\
$\pm$            & 0.05 & 0.02 & 0.02 & 0.03 & 0.08 \\
\hline
$\Delta$m        & 0.88 & 0.80 & 0.70 & 0.66 & 0.63 \\
$\pm$            & 0.10 & 0.04 & 0.04 & 0.06 & 0.15 \\
\hline
\end{tabular}
\label{EDR}
\end{table}

\section[]{Identification of the lenses}

If these systems of paired QSOs are indeed lensed QSOs, then the lenses
should be identifiable as clusters of galaxies.  As described
in section 2, one
of the possible lensed systems, J$0223-2906$, appears associated with a
galaxy cluster at $z = 0.11$.  Ten galaxies in the vicinity have  
redshifts, measured by the 2dF Galaxy Redshift Survey, 
that are close to this value, with a velocity
dispersion of about 600~km~sec$^{-1}$ (Fig.~\ref{fchart}).  
Such a velocity dispersion would
however produce an image splitting of only about $20''$ for an isothermal
sphere lens, so if this cluster
is the lens then either an anisotropic or a more centrally
concentrated mass distribution is needed.  The total
mass required to produce the observed image splitting is 
$\sim 6 \times 10^{14}$M\sun.  There is also a weak ROSAT-detected X-ray source
close to the position of the two bright galaxies which lie along the line
joining the two QSOs, although the origin of the emission is not known.
At present it is not clear if sufficient mass is
present in this cluster to produce the observed image splitting.
One other system, J0150$-$2921, 
may also have a low-redshift cluster
associated with it, again based on redshift measurements from the 
2dF Galaxy Redshift Survey,
although in this case only three redshifts are
currently known (Fig.~\ref{fchart}).

In order to estimate the expected distribution of lens redshifts we
consider the calculations presented by \citet{lopes03} in which the
lensing clusters are assumed to have spherically-symmetric \citet{NFW} 
mass profiles, and
to have an abundance predicted by the \citet{st99} evolving mass function.
Here we assume a cold dark matter power spectrum 
\citep{bardeen} modified to include an approximate correction for baryons
\citep{sug95} in a $\Lambda$CDM cosmology with $\Omega_{\rm M} = 0.3$
and $\Omega_{\Lambda} = 0.7$ and a power-spectrum normalisation
given by $\sigma_8 = 0.9$. In this model, it is found that the distribution
of lens redshifts is not sensitive to the choice of cosmological parameters,
although the abundance of lensing systems is. Magnification bias is calculated
as discussed by \citet{lopes03}. 

Within this model, it is clear that a lensing cluster is expect to have
$z_{\rm LENS} < 1$, and for the widest separations the model predicts that
any lens should have $z_{\rm LENS} < 0.5$.  However, the total probability
predicted for $200''$ lensed systems is so low that such systems are expected
only if there are significant departures from the assumptions of the model,
in which case the lens redshift distribution would need to be recalculated.

\begin{figure}
\resizebox{75mm}{!}{
\rotatebox{-90}{
\includegraphics{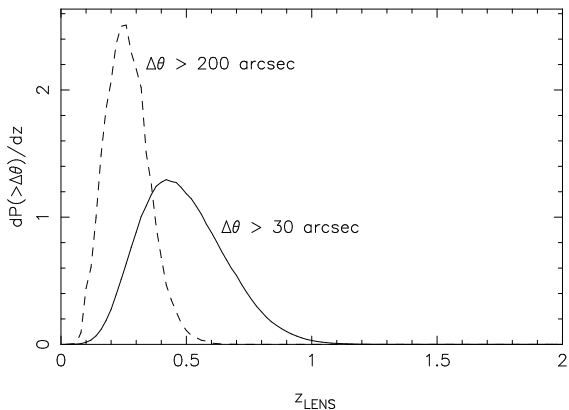}}}
\caption{
The {\em a posteriori} probability distribution of redshift,
$z_{\rm LENS}$, for a lensing
cluster producing multiple images of a
QSO at $z=2$ with separations either $> 200''$ or $> 30''$, 
assuming the $\Lambda$CDM cosmology and 
cluster lens model described in the text.
\label{fig4}
}
\end{figure}

\section{Comparison with models}

At this stage it cannot be regarded as conclusive that these candidate
systems are indeed lensed, and we also do not yet have an accurate
assessment of the completeness of the lensing search.  Nonetheless it
is interesting to compare the results presented here with the
expectations of lensing models.  \citet{li02} and \citet{lopes03} have
calculated the expected distribution of angular separations arising
from strong lensing by massive clusters of galaxies in the standard
cosmological framework that assumes structure grows hierarchically
from initial Gaussian perturbations in a universe dominated by cold 
dark matter and dark energy.  With cosmological parameters as
determined from the WMAP observations \citep{spergel03} the expected
number of systems with $\Delta\theta > 10''$ in the 2QZ survey of $2
\times 10^4$ QSOs is about 0.5 \citep{lopes03}, so it is possible,
although not highly likely, that the 2QZ survey contains
wide-separation lensed systems.  However, the distribution is
expected to be dominated by angular separations towards the lower end
of the range, and in this standard model the expected number of
systems with $\Delta\theta > 30''$ is only $\sim 0.1$ \citep{lopes03}.
One possible explanation is that the widest-separation systems presented
here are not in fact lensed, despite the similarity of the redshifts
and spectra.  A second is that the values of cosmological and cluster
model parameters are not as assumed in calculating the above numbers:
key parameters are $\sigma_8$, $w$, $\Omega_M$ and also parameters of
the model such as the dependence of cluster halo concentration on
mass, redshift and cosmology and the cosmic scatter in halo
concentration (for details see \citealt{lopes03}).  Or it may be that
the models do not yet comprise the full complexity required to model
accurately the statistics:  possible deficiencies include the
assumption of the \citet{NFW} mass profile (see also
\citealt{oguri03}) and the assumption of a smooth distribution of
mass.  Other departures from the standard cosmological model, such as
the existence of non-Gaussian fluctuations, could also modify the
lensing probabilities.

For these reasons we believe it is imperative to test carry out
observational tests of the candidate lensed systems presented here in
order to test these models, rather than using the predictions of the
models to argue against the interpretation of these systems as being lensed.

\section{Comparison with previous observations}

Previous searches have been undertaken for lensed QSOs with comparable
separations, and it is interesting to investigate whether the detection
of any systems in 2QZ would be inconsistent with those searches.
  
\citet{ofek01, ofek02} searched a sample of size $\sim 8000$
radio-loud QSOs without finding any evidence for multiple images in
the separation range $5'' < \Delta\theta < 30''$.
\citet{phillips01b} examined a complete sample of 11,670
flat-spectrum radio sources for possible multiple images separated 
in angle by $6'' < \Delta\theta < 15''$ and found no lensed systems.
An incomplete sample of $\sim 15,000$ radio sources yielded one possible
lensed system.  
The first step in comparing these results with those from the 2QZ
survey is to compare sample selection.  

The range of angular separations probed by the radio surveys is generally
smaller than considered here, although had any 2QZ pairs with
$\Delta\theta > 10''$ been lens candidates they should have been
identified as such.  For comparison purposes we shall use the
cluster-lensing model to compare surveys covering different ranges in
separation, although we have argued above that the overall lensing
probabilities predicted by such models may be too low to explain the
existence of more than one lensed system in the 2QZ survey.  
The model of \citet{lopes03} with parameters $\Omega_M =
0.3$, $\Omega_{\Lambda} = 0.7$, $\sigma_8 = 0.9$ and $z_{\rm QSO} =
1.5$ predicts relative lensing probabilities of $1$, $1.7$
and $1.3$ for the separation ranges $6-15''$, $5-30''$
and $>10''$ respectively, assuming the same magnification bias in each
case.  If the models are indeed deficient in some way, or if the
values
of parameters are different from that assumed, these relative
probabilities will be different, but they suffice for the discussion
in this section.

In fact, because of the double radio and optical selection of
the FIRST survey \citep{ofek01} the magnification bias could be
a factor 5 smaller for that survey compared with the JVAS/CLASS or 2QZ
surveys (\citealt{ofek02, phillips01b, lopes03}), 
so the overall expected relative
lensing probabilities in the three surveys are $0.2$, $1.7$ and $1.3$ 
for FIRST, JVAS/CLASS and 2QZ respectively.  

We also need to make some assumption about sample completeness and to
multiply the probabilities by the sample sizes. Here
we shall adopt a conservative view of
the JVAS/CLASS result and say that out of 15,000 QSOs no lensed systems
were found,
and we adopt an assumed completeness of 50\% for the 2QZ survey.  At
this stage this figure is illustrative only.  With these assumptions,
the predicted relative numbers of systems in each survey are
0.1, 1.6 and 1.0 for the FIRST, JVAS/CLASS and 2QZ surveys respectively.

Now suppose the entire dataset contains one lensed system: the {\em a
posteriori} probability of that system being found in the 2QZ survey is
37\%.  However, if there are more systems in the 2QZ survey, the
probability of there being none in the other surveys falls to
13, 5, 2 \% for 2, 3 or 4 2QZ systems.  Thus, if the surveys are
indeed comparable in this way, the existence of two lensed systems
within the 2QZ survey would not be inconsistent with the other
surveys, but a larger number could be.

Finally, however, we note that another possibly significant factor in calculating the
expected lensing probabilities is the
redshift distribution of the source population.  \citet{wambs03},
\citet{oguri03} and \citet{lopes03} have demonstrated that cluster lensing
probability strongly increases with the mean redshift of the source 
population, and also depends to some extent on the distribution of
source redshifts about the mean.  The redshift distribution is not
well-known for the CLASS and FIRST surveys, but appears likely to have
a median value in the range $1.0 < z < 1.4$ 
(\citealt{ofek02,phillips01a,phillips01b}).  The 2QZ median redshift is
1.5.  Lensing probability increases by a factor about 3 as the source
redshift is increased from $z=1$ to $z=1.5$ \citep{lopes03}, 
so if the typical redshift
of sources in the radio surveys is even slightly lower than that of QSOs
in the 2QZ survey this would also lower the relative probability of 
discovering lensed systems in the radio surveys.  

Our overall conclusion from this comparison is that, observationally,
the results from the 2QZ and radio surveys may not be incompatible,
but clearly further work needs to be done both to establish more
accurately the redshift distribution of the JVAS/CLASS sources and to
establish the reality of the 2QZ candidate systems.

\section{Conclusions}

Evidence has been presented for the discovery of the first lensed QSOs 
with multiple images separated on arcminute scales.  If these are
genuine lensed systems then we should expect the associated lenses 
to be observable as massive clusters of galaxies.  Further evidence
needs to be collected to establish that these are indeed lensed, however.
One such piece of evidence that is independent of assumptions of the
nature of the lens would be the existence of common Ly-$\alpha$ forest
absorption in candidate lensed QSO spectra.  The existence of giant
arcs associated with a system would also provide strong evidence for
cluster lensing, especially if the host galaxies of the QSOs could
be detected and shown to be distorted by lensing.  Detection of the
lens itself, either by direct imaging or weak-lensing, would also
provide convincing evidence.
  
Consideration of the expected distribution of angular separations 
indicates that, although a number of the identified systems may be
lensed, a number of those with the greatest separations may
not be.  Confirmation or discounting of the widest separated systems 
is particularly important as any such systems would challenge our
current understanding of lensing by galaxy clusters.

Among the examples found, one candidate lensed system has been
found to display broad CIV absorption in one image and not the other,
which could be due to the formation of such absorption systems on 
timescales of tens of years.  Further searches for wide-separation lenses
may be partially incomplete because such absorption systems would 
affect optical colours, spectra and the flux in other wavebands, including
possibly the X-ray band.  Broad-line region variability may also 
render search methods
based on colour and/or spectroscopic comparison partially incomplete
unless generous selection criteria are allowed. 
Finally, it may be that optical searches are biased against
lenses with separations less than a few tens of arcsec because of
obscuration in galaxies associated with the cluster lens.

\vspace*{5mm}

\noindent
{\large \bf ACKNOWLEDGMENTS}\\

The 2dF QSO Survey was based on observations made with the U.K.~Schmidt
and Anglo-Australian Telescopes.  We thank all the present and former
staff of the Anglo-Australian Observatory for their work in
building and operating the 2dF facility.  Follow-up spectroscopic
observations were carried out at the Anglo-Australian Telescope and
at the William Herschel Telescope.  The William Herschel Telescope is
operated on the island of La Palma by the Isaac Newton Group in the
Spanish Observatorio del Roque de los Muchachos of the Instituto de
Astrofisica de Canarias.  
We acknowledge the use of the Sloan Digital Sky Survey archive for
the photometry presented in Table~\ref{EDR}.
We are also grateful to the 2dF Galaxy Redshift Survey team for
providing redshifts of galaxies in the fields shown here, in
advance of publication, and would like to thank Raylee Stathakis
for useful discussions and Aidan Crook for assistance with observing
preparation. AML acknowledges the support of the 
Portuguese Funda\c{c}\~{a}o para a Ci\^{e}ncia e a Tecnologia.

\label{lastpage}

\end{document}